\documentclass{elsart}

\usepackage{natbib}   
\usepackage[dvips]{graphicx}
\usepackage{amssymb}

\begin{document}
\begin{frontmatter}

\title{Fine-tuning in the rotation curves of spirals and the uniqueness of their mass modelling.}
\author{Chiara Tonini, Paolo Salucci}

\ead{tonini@sissa.it}
\ead{salucci@sissa.it}

\address{International School for Advanced Studies (SISSA), via Beirut 2-4, 34014 Trieste, Italy}

\begin{abstract}
We construct a generic mass model of a disk galaxy by combining the usual two mass components, an exponential thin stellar disk and a dark halo for which we alternatively assume one of the two most common mass distributions. We construct its generic rotation curve, with two free parameters linked to its shape and one to its amplitude, namely the halo characteristic scale lenght, the fractional amount of Dark Matter and the total mass at one disk scale lenght.\\
We find that there is no need of a fine-tuning in the above parameters to produce smooth, featureless rotation curves like the ones actually observed. The rotation curves obtained from different mass distributions are significanlty affected by the variation of the parameters, and show substantial difference in their slope; however, they do not feature irregularities marking the transition from a disk-dominated to a halo-dominated regime.\\
On the other hand we show, by direct analysis of a large number of plausible and different mass models, that an observed rotation curve fulfilling well defined quality requirements can be uniquely and properly decomposed in its dark and luminous components; through a suitable mass modelling together with a maximum-likelihood fitting method, in fact, the disentangling of the mass components is accomplished with a very high resolution.
\end{abstract}

\begin{keyword}
galaxies: rotation curves --- galaxies: Dark Matter halos --- galaxies: mass components
\end{keyword}

\end{frontmatter}

\section{Introduction}

The rotation curves of spiral galaxies are observed to be smooth, featureless, out to several disk scale lenghts: they are essentially linear functions of the galactocentric distance, with a hint of flattening at the outermost radii (Persic, Salucci \& Stel, 1996; see Appendix A). Rotation curves are determined by the full distribution of the mass of a galaxy, including its luminous component (the disk) and its dark component (the halo). While the radial distribution of the luminous matter is observed for spiral disks, the disk mass itself is unknown. Regarding the Dark Matter, both the amount and the distribution are unknown, and since these physical quantities affect the rotation velocity profile in a different way at different radii, the study of the rotation curves becomes a very important tool to investigate the structure of galaxies.\\
There is sometimes the idea that the smoothness of the observed rotation curves implies a fine-tuning of the halo/disk structural parameters. Related to it, although logically orthogonal, there is the view that, due to this smoothness, it is impossible to disentangle the dark and luminous components of a RC, because very different mass models are thought to ``produce'' the same rotation curve\footnote{Due to the results of this paper, we find no reason to explicitly identify the supporters of these ideas.}. The aim of this paper is to show that both these ideas are incorrect, \textit{i.e.} that \textit{a)} the characteristics of the observed RCs do not imply the existence of a fine-tuning in the halo and the disk structural parameters, but, on the contrary, these parameters take all the physically possible values and still produce smooth RCs, and \textit{b)} ``physically'' different mass models yield detectably different RCs, provided that the observational data fulfill well defined high-quality requirements. \\
We model the rotation curves of a galaxy from the distribution of the luminous and Dark Matter. We do not aim at the determination of the true Dark Matter density profile, issue that has to be dealt with by a different approach. Instead we assume the most popular density profiles, and we investigate the characteristics of the corresponding RCs. In detail, we study a pseudo-isothermal (PI) halo profile, suggested by observations (Persic et al. 1996), and a Navarro, Frenk \& White (NFW) halo profile (Navarro, Frenk, \& White, 1997) emerged in $CDM$ N-body simulations.  \\
In section 2 we present the rotation curves derived for a PI halo profile (2.1), and for a NFW halo profile (2.2), and we discuss their characteristics. In section 3 we investigate the possibility of uniquely deriving, from an  high-quality rotation curve, the actual mass model. Finally, in section 4 we  summarize the results. \\   

\section{Generic mass models produce smooth, featureless Rotation Curves}
The total mass of a disk galaxy can be divided into two independent components, namely the luminous disk and the Dark Matter halo (neglecting the bulge is irrelevant in our study). These components contribute separately to the total rotation velocity:
\begin{equation} 
V^2(\tilde{r}) = V^2_d(\tilde{r}) + V^2_h(\tilde{r}) 
\end{equation}
For the luminous matter, we consider the stellar disk, whose surface density profile is found as:
\begin{equation}
\Sigma(r) = \Sigma_0 e^{-R/R_D}
\end{equation}
where $\Sigma_0$ is the central surface density, and $R_D$ is the disk scale lenght\footnote{It's worth noticing the observational relationship between the circular rotation velocity ($V_3 \equiv V(3R_D)$) and $R_D$:  $R_D \simeq 5 \ \left( \frac{ V_3 }{ 200 } \right)  kpc$ (Persic et al. 1996).
}.
The disk component of the velocity is (Freeman, 1970):
\begin{equation} 
V^2_d(\tilde{r}) = \frac{ G M_D }{ R_D } \ \nu(\tilde{r})
\end{equation}
in which $M_D$ is the disk mass and $\nu(\tilde{r})=\tilde{r}^2/2 \cdot (I_0K_0 - I_1K_1 )|_{\tilde{r}/2}$, with $I_0K_0$ and $I_1K_1$ being the Bessel functions of order 0 and 1, evaluated at $\tilde{r}/2$. \\
\begin{figure}
\begin{center}
\includegraphics[width=8cm,height=8cm]{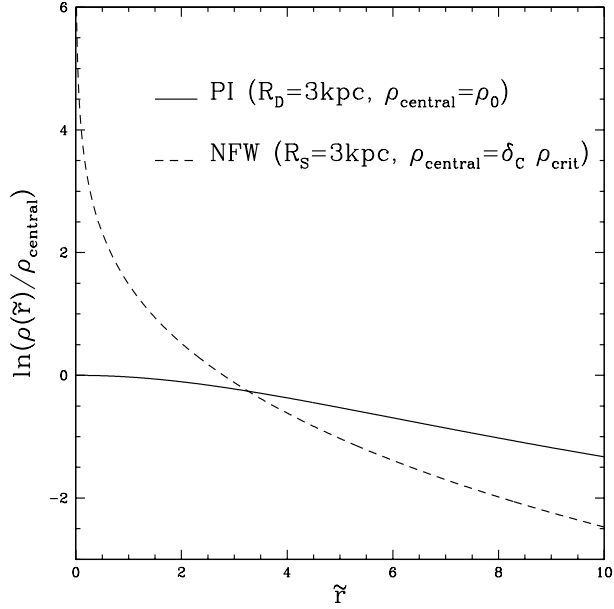}
\caption{The normalized PI and NFW halo density profiles.}
\label{profili}
\end{center}
\end{figure}
For the DM halo, we investigate both the PI profile and the NFW profile, (see figure (1) for comparison), with the halo velocity component being:
\begin{equation}
V^2_h(\tilde{r}) = \frac{G M_H(\tilde{r})}{\tilde{r} \ R_D}
\end{equation}
where $M_H(\tilde{r})$ is the dark halo mass inside the radius $\tilde{r}$.\\
The contribution of the luminous matter to the total rotation velocity is determined by two parameters: the disk scale lenght $R_D$, measured in each galaxy, and the total mass $M_D$, that can be obtained, at least in principle, from the disk luminosity, but since its determination is affected by large uncertainties, it is usually considered as an unknown quantity. The dark component of the rotation velocity has two free parameters, the central halo density and the halo characteristic scale lenght, that fully determine the DM distribution.\\

\subsection{PI halos}
The DM density distribution of \it{pseudo-isothermal} \rm (PI) halos is given by: 
\begin{equation}          
\rho(r) = \frac{ \rho_0 }{ 1 + (r/ R_C )^2 } 
\end{equation}
For convenience we define: $\tilde{r} \equiv r/R_D$ as the galactocentric distance, and $\tilde{R}_C  \equiv  R_C/R_D$ as the core radius of the halo, both in units of the disk scale lenght. We take $\tilde{R}_C$ as a free parameter and we consider its possible variation range between 0 and infinity. $\tilde{R}_C < 1$ implies a mass distribution very similar to that of the NFW profile (see subsection 2.2). On the other hand, cases with $\tilde{R}_C > 3$ are scarcely distinguishable from one another, since most of the observational data do not reach regions of disks beyond $\tilde{r}=3$.\\ 
The mass of the dark halo as a function of the normalized radius is obtained as: 
\begin{equation}
M_H(\tilde{r})=4\pi G \rho_0 \ \tilde{R}_C^3 R_D^3 \left[\frac{\tilde{r}}{\tilde{R}_C} - tan^{-1}\left(\frac{\tilde{r}}{\tilde{R}_C}\right)\right]
\end{equation}
We define:
\begin{equation}
\lambda (\tilde{r}) \equiv \frac{\tilde{R}_C^3}{\tilde{r}} \ \left[\frac{\tilde{r}}{\tilde{R}_C} - tan^{-1}\left(\frac{\tilde{r}}{\tilde{R}_C}\right)\right]
\end{equation}
The total rotation velocity of the galaxy can therefore be written as:
\begin{equation}
V^2(\tilde{r}) = \frac{ G M_D }{ R_D } \ \nu(\tilde{r}) + \frac{G 4\pi \rho_0 \ R_D^3}{R_D} \ \lambda(\tilde{r})
\end{equation}
It is useful to perform the following transformation: 
\[\alpha \equiv \frac{G M_D}{R_D \ V^2_1}, \hspace{0.7cm} \beta \equiv \frac{4 \pi G \rho_0 \ R_D^3}{R_D \ V^2_1}, \hspace{0.7cm} \lambda_1 \equiv \tilde{R}_C^3 \ \left[ \frac{1}{\tilde{R}_C} - tan^{-1}\left(\frac{1}{\tilde{R}_C} \right)\right]\]
The parameters $\alpha$ and $\beta$ are directly proportional to the fraction, at $R_D$, of the luminous and dark mass respectively. We define $V^2_1 \equiv V^2(\tilde{r}=1))$. 
The radial profile of the normalized rotation velocity is then given by: 
\begin{equation}
\frac{V^2(\tilde{r})}{V^2_1} = \alpha \ \nu(\tilde{r}) + \beta \ \lambda(\tilde{r})
\end{equation}
$\alpha$, $\beta$ and $\lambda$ determine the shape of the curve, while $V^2_1$ determines its amplitude. Let's stress that, as concerning the results of this section, the actual value of the constant $V_1$ is irrelevant since, as a normalization parameter, it doesn't affect the RC shape. From (9) it follows that:
\begin{equation}
\alpha \ \nu_1 + \beta \ \lambda_1 = 1
\end{equation}
where $\nu_1=\nu (\tilde{r}=1)$. We  notice that $\alpha \nu_1$ and $\beta \lambda_1$ represent the fractional contributions to the total (normalized) velocity at $R_D$ due to the luminous and Dark Matter respectively; for simplicity we define $f_{DM} \equiv \beta \lambda_1$. \\
We now rewrite the radial profile of the normalizated rotation velocity as
\begin{equation}
\frac{V^2(\tilde{r})}{V^2_1} = (1-f_{DM})\ \frac{\nu(\tilde{r})}{\nu_1} + f_{DM} \ \frac{\lambda(\tilde{r})}{\lambda_1}
\end{equation}
under the following ranges of variations: 
\[0 \leq f_{DM} \leq 1  \hspace{1cm} 0 \leq \tilde{R}_C \leq 3\]
In this way, we split the original three free parameters into a 2$+$1 combination: 2 determine the RC profile and 1 sets the RC amplitude. We allow the free parameters related with the profile to vary over a very wide range of values, and we plot the corresponding curves as functions of radius $\tilde{r}$.\\
\begin{figure*}
\begin{center}
\includegraphics[width=14cm,height=9cm]{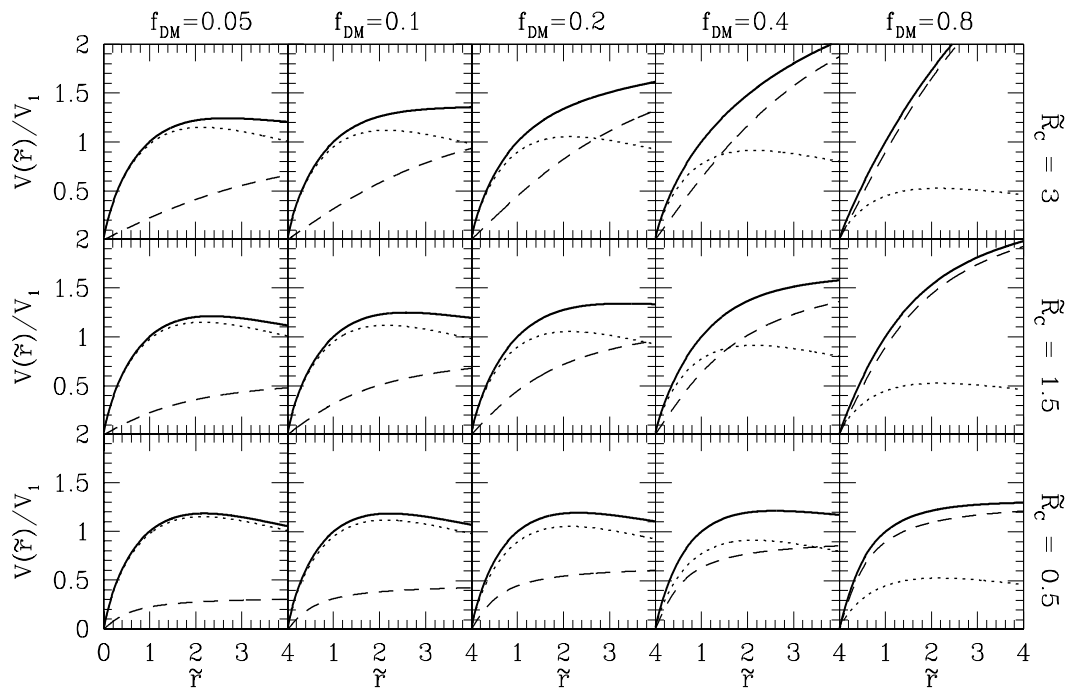}
\caption{PI rotation curves, for different $f_{DM}$ and normalized $\tilde{R}_C$. The curves represent the total rotation velocity (thick line), the disk-component velocity (dotted lines) and the halo-component velocity (dashed lines).}
\label{iso15}
\end{center}
\end{figure*}
\begin{figure*}
\begin{center}
\includegraphics[width=10cm,height=10cm]{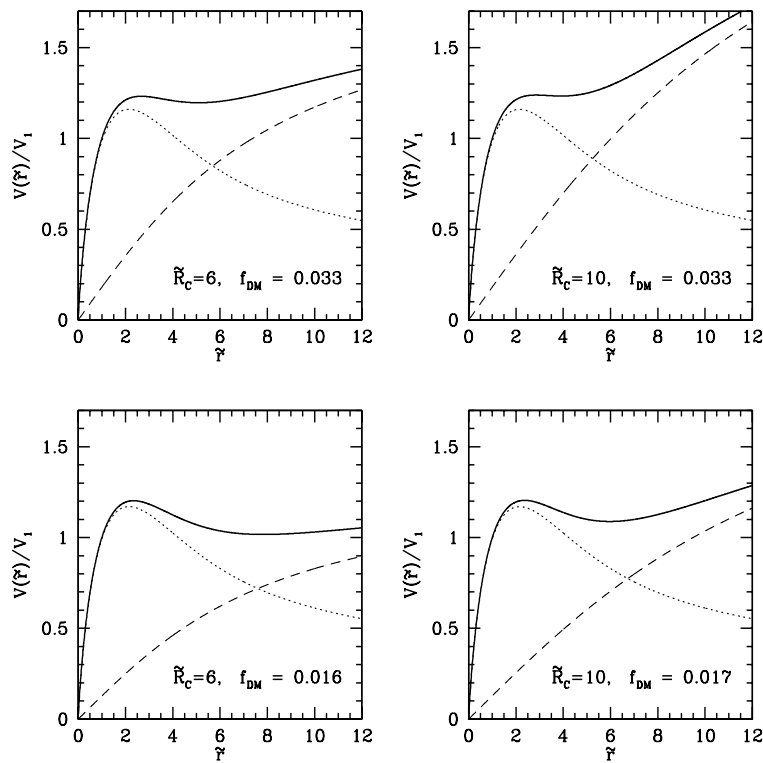}
\caption{PI rotation curves with a second-order anomaly, indicating a sharp transition between a disk-dominated to a halo dominated region.}
\label{isof}
\end{center}
\end{figure*}
In figure (2) we present typical rotation curves we obtain for a variety of cases, including those with almost no Dark Matter in the central regions and those strongly halo-dominated, as well as those with large and small core radii. The curves are plotted for $0 \leq \tilde{r} \leq 4$, covering the radial range of the observations. \\
We realize that the variations of the parameters affect only the RCs' slopes: as $f_{DM}$ increases, the curves become progressively more halo-dominated and therefore steeper, and the effect is enhanced for increasing $\tilde{R}_C$. However, as the parameters vary, there are no peculiar features in the RCs, no flexes indicating a discontinuity between a disk-dominated and a halo-dominated regime. In fact, the transition between these two regions appears smooth, and does not cause irregularities in the curves, at least in the zones actually mapped by data.
No fine-tuning in the amount of the Dark Matter or in its density distribution parameters is required to obtain a smooth and featureless RC, like, for instance, those of the URC (Persic et al. 1996; see also Appendix A). \\ 
Notice however that we can build a few peculiar models in which the transition between a disk-dominated to a halo-dominated region makes itself visible in the RC profile, locally modifying the RC shape in a significative way. In figure (3) we show four of these cases. Anyway, they are very extreme, in that these galaxies have unrealistically low amounts of Dark Matter, $f_{DM} \leq 3 \% $, and very large core radii, $\tilde{R}_C \ge 6$. Moreover, the flex in the curve, although quite significant, appears at galactocentric distances that are beyond the usual outermost radius reached by observational data. It is worth to point out that these peculiar RCs have the same profile often thought to be the standard outcome of the luminous and Dark Matter distributions in disk galaxies. Instead, to be detected in the RCs, these features require a careful setting of the free parameters. The general properties of the dark and luminous matter that we observe in spirals make sure that, if a fine-tuning is needed to reproduce an observed RC, it is to feature irregularities, not smoothness. \\

\subsection{NFW halos}
We now adopt the Dark Matter distribution of NFW halos (Navarro et al. 1997): 
\begin{equation}
\frac{\rho(r)}{\rho_{crit}} = \frac{\delta_C}{(r/R_S) \ ( 1 + r/R_S )^2}
\end{equation}
where the density is normalized to the critical density of the Universe, and $\delta_C$ is the central density contrast (with respect to the external background). $R_S$ is the halo characteristic lenght scale, and again we define: $\tilde{R}_S \equiv R_S/R_D$ and $\tilde{r} \equiv r/R_D$. The mass of the dark halo inside a given radius is now: 
\begin{equation}
M_H(\tilde{r}) = 4\pi \rho_{crit}\delta_C\tilde{R}_S^3 R_D^3 \left[ ln \left( 1+\frac{\tilde{r}}{\tilde{R}_S} \right) - \frac{1}{(1+\tilde{R}_S/ \tilde{r} )}\right]
\end{equation}
After defining, in replacement of equation (7):
\begin{equation}
\omega(\tilde{r}) \equiv \frac{\tilde{R}^3_S}{\tilde{r}} \ \left[ ln \left( 1+\frac{\tilde{r}}{ \tilde{R}_S} \right) - \frac{1}{(1+\tilde{R}_S/ \tilde{r} )} \right]
\end{equation}
and using equations (1), (3) and (4), we write the expression for the total rotation velocity:
\begin{equation}
V^2(\tilde{r}) = \frac{ G M_D }{ R_D } \ \nu(\tilde{r}) + \frac{G 4\pi \rho_{crit} \delta_C \ R_D^3}{R_D} \ \omega(\tilde{r})
\end{equation}
Similarly to the previous subsection, we set:
\[\alpha \equiv \frac{G M_D}{R_D \ V^2_1}, \hspace{0.5cm} \beta \equiv \frac{4 \pi G \rho_{crit} \ \delta_C \ R_D^3}{R_D \ V^2_1}, \hspace{0.5cm} \omega_1 \equiv \tilde{R}_S^3 \left[ ln \left(1+\frac{1}{\tilde{R}_S} \right) - \frac{1}{1+\tilde{R}_S} \right]\]
and we get the normalized total velocity profile: 
\begin{equation}
\frac{V^2(\tilde{r})}{V^2_1} = \alpha \ \nu(\tilde{r}) + \beta \ \omega(\tilde{r})
\end{equation}
with the constraint given by equation (10); finally, we obtain the parametrized expression for the rotation curve:
\begin{equation}
\frac{V^2(\tilde{r})}{V^2_1} = (1-f_{DM}) \ \frac{\nu(\tilde{r})}{\nu_1} + f_{DM} \ \frac{\omega(r)}{\omega_1}
\end{equation}
\begin{figure*}
\begin{center}
\includegraphics[width=14cm,height=9cm]{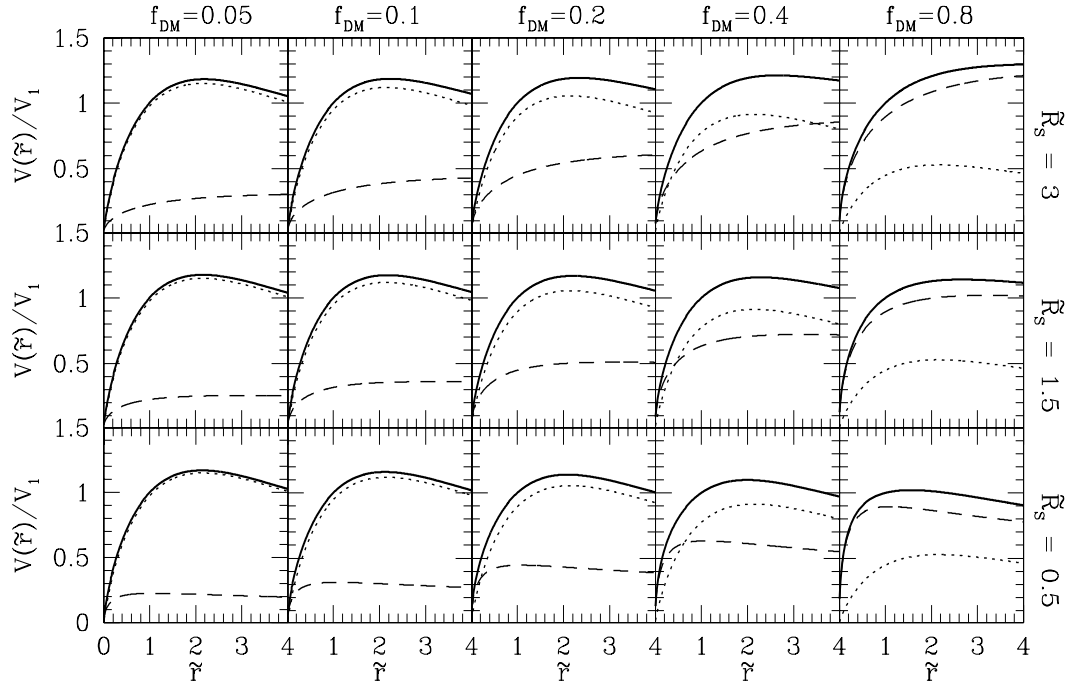}
\caption{NFW rotation curves, for different $f_{DM}$ and normalized $\tilde{R}_C$.}
\label{NFW15}
\end{center}
\end{figure*}
Again, the two parameters $f_{DM}$ and $\tilde{R}_S$ completely characterize the rotation curve profile. In figure (4) we show the NFW RCs for the same values of the parameters as the PI case, and we find a very similar behaviour: in general, $NFW+disk$ curves are even smoother and flatter than the $PI+disk$ ones; also in this case the emerging RCs have no second-order irregularities associated to a luminous-to-dark regime transition, at least in the regions contained inside 4 disk scale lenghts.  \\

\smallskip
In the region contained inside the stellar disk edge, general PI and NFW mass models predict, for all values of the model free parameters,  RCs that are smooth and featureless, \textit{i.e.} with no flexes and inversion of the first derivative.

\section{Disentangling a high-quality RC}
RCs are smooth and featureless when considered at small scales, and they show no individual charecteristics, apart from a different value of their ``large-scale'' slope. This is evident in the model RCs of the previous section and in the observational scenario provided by the URC.\\
The further question is: how well does a rotation curve yield the true mass model of a galaxy? Moreover, is the information we obtain from the observational data unique and model-independent? In other words, can we distinguish between different Dark Matter amounts and distributions by analyzing a rotation curve?
In this section we will answer these questions for the PI case, which presents a number of difficulties, since it is partially degenerate with respect to its structural parameters $f_{DM}$ and $\tilde{R}_C$. The NFW case will not be considered here in that, having one free parameter less than the PI case and a known radial behaviour of the velocity profile, any result we find in our investigation for the PI will apply also to the $CDM$ profile.\\
\begin{figure}
\begin{center}
\includegraphics[scale=0.7]{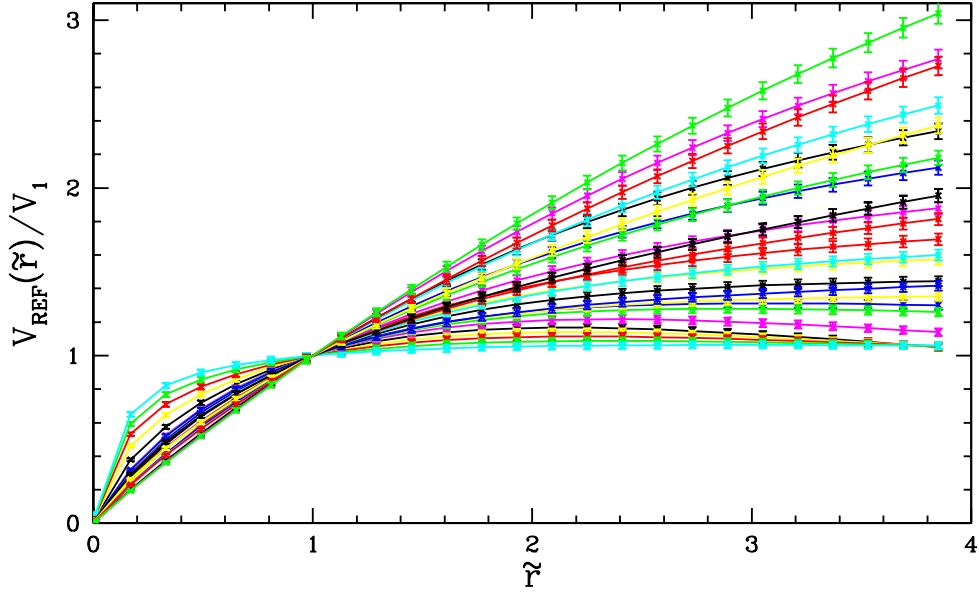}
\caption{The 25 reference ``observed'' rotation curves.}
\label{reference}
\end{center}
\end{figure}
Let's assume then that galaxies are composed by an exponential thin disk surrounded by a PI dark halo, and estimate the precision with which we can resolve the luminous and the Dark Matter distribution from the rotation curve. Notice that the Universal Rotation Curve and the individual RCs are well fitted by a $PI+disk$ model (Donato, Gentile \& Salucci 2004; Gentile et al. 2004). \\
We build a family of 25 reference curves that simulate ``observed'' RCs of real galaxies, each one consisting of 25 data points, evenly spread between 0 and 4 disk scale lenghts, and is characterized by a set of two parameters in the ensemble: $f_{DM} = \{0.1,0.3,0.5,0.7,0.9\}$ $\otimes$ $\tilde{R}_C =\{0.1,1.0,2.0,3.0,4.0\}$. To each point of these curves we assign ``observational'' errors of the order to those of high-quality published RCs (Persic \& Salucci 1995): $\epsilon_V = 0.02$ is the averaged uncertainty in $V(\tilde{r})$, and $\epsilon_D = 0.05$ in the RC slope. These curves are shown in figure (5). \\
We investigate each of the ``observed'' RCs with a mass model, fitting the data to obtain their structural parameters. We remind that a successful strategy is accomplished by means of three suitable free parameters, namely the dark halo core radius $\tilde{R}_C$, the relative contribution of the Dark Matter to the total velocity $f_{DM}$ and the absolute value of the rotation velocity at a distance from the centre of one disk scale lenght $V_1$. Of these parameters, two affect the shape of the RC, and one, $V_1$, affects its amplitude. If the observed RC is of high quality, then it's possible to set the value of $V_1$ directly from the data, reducing the number of free parameters. In this case,
\begin{equation}
V_1 \equiv V(1) \pm \mathcal{O} (10^{-2}) 
\end{equation}
Let us caution that this cannot always be done: we define a high-quality RC as a set of kinematical data for which such a parameter reduction is possible. \\
Therefore, the reference ``observed'' RCs are fitted with the following mass model (see equation 11): 
\begin{equation}
V^2_{mod}(\tilde{r})=V^2(1) \ [(1-f_{DM}) \ \gamma_L(\tilde{r}) + f_{DM} \ \gamma_{DM}(\tilde{r};\tilde{R}_C)]
\end{equation}
where $\gamma_L(\tilde{r}) \equiv \nu(\tilde{r})/\nu_1$ is known, and $\gamma_{DM}(\tilde{r};\tilde{R}_C) \equiv \lambda(\tilde{r};\tilde{R}_C)/\lambda_1(\tilde{R}_C)$. \\
\begin{figure}
\begin{center}
\includegraphics[width=15cm,height=11cm]{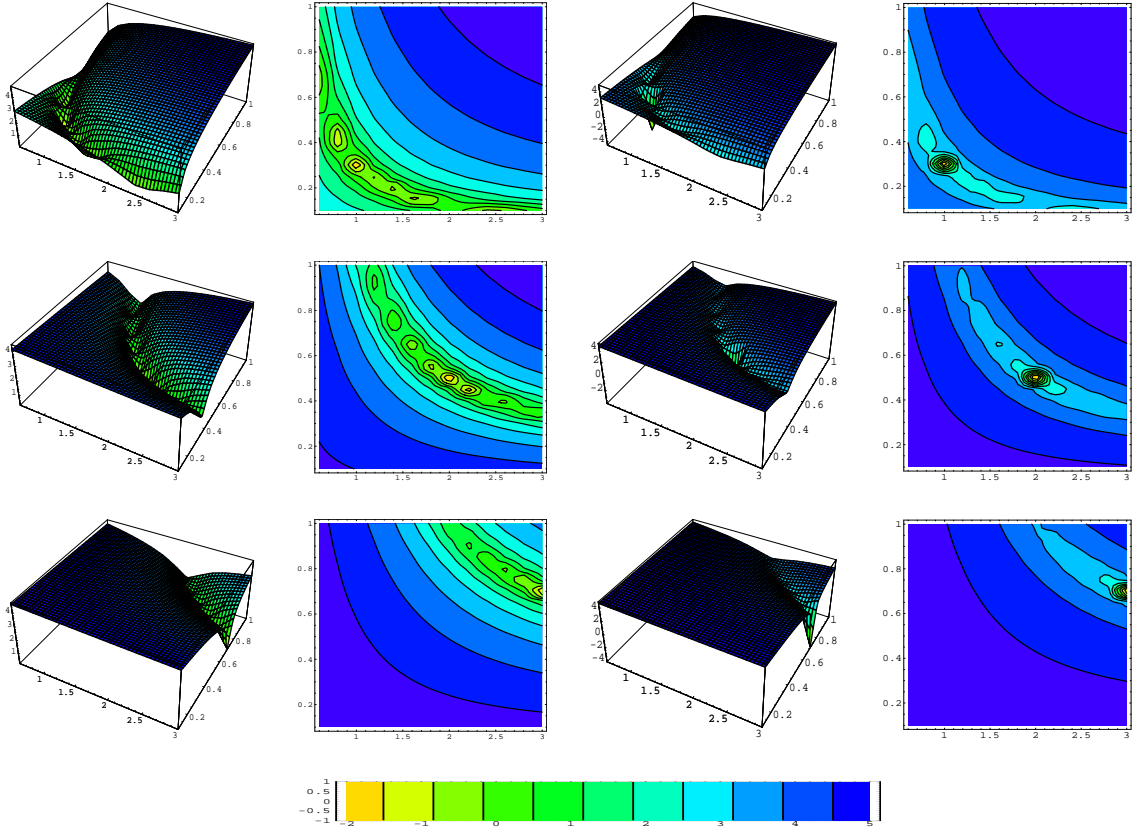}
\caption{Left: $\chi^2(V)$ for 3 reference curves, as a function of the model parameters: $\tilde{R}_C \in [0.2,4]$ (x-axis) and $f_{DM} \in [0.05,1]$ (y-axis). The ``true'' parameters are: ($\tilde{R}_C=1.0,$$f_{DM}=0.3$) (top panels), ($\tilde{R}_C=2,f_{DM}=0.5$) (middle panels), ($\tilde{R}_C=3,f_{DM}=0.7$) (bottom panels). Right: $\chi^2_{TOT}$ for the same cases.}
\label{vel}
\end{center}
\end{figure}
The measure of the distance between a model curve and an ``observed'' RC is given by a likelihood parameter, that we define as the sum of the $\chi^2$ computed on the velocity and on the RC slope:
\begin{equation}
\chi^2_{TOT}=\chi^2[V]+\chi^2 [D]
\end{equation}
where the terms on the right-hand side are:
\begin{equation}
\chi^2_V([f_{DM},\tilde{R}_C]_{mod})= \frac{1}{\epsilon_V^2} \ \sum_{n=1}^{25} \left(V_{mod}(\tilde{r}_n;[f_{DM}, \tilde{R}_C)]_{mod})-V_{obs}(\tilde{r}_n)\right)^2
\end{equation}
\[\chi^2_D([f_{DM},\tilde{R}_C]_{mod})=\]
\begin{eqnarray}
 \hspace{1cm} \frac{1}{\epsilon_D^2} \ \sum_{n=1}^{25} \left(\frac{\tilde{r}_n}{V_{mod}} \ \frac{dV_{mod}(\tilde{r}_n;[f_{DM}, \tilde{R}_C)]_{mod})}{d\tilde{r}}-\frac{\tilde{r}_n}{V_{obs}} \ \frac{dV_{obs}(\tilde{r}_n)}{d\tilde{r}}\right)^2
\end{eqnarray}
We denote with $V_{mod}$ and $V_{obs}$ the velocities of the model curves and of the reference curves respectively.\\
The need of considering the contribution of the gradient of the velocity in the computation of the likelihood parameter is well explained by the examples shown in figure (6). In the left panels we show the value of the $\chi^2_V$ only, as a function of the values of the model parameters, for three different reference curves. The $1\sigma$ contour of $\chi^2_V$ includes also values of the parameters far from the ``true'' ones, unveiling a certain inability of the model to disentangle the mass components. With the information given by the gradient of the circular velocity and by the $\chi^2_{TOT}$ as defined in equation (20), the degeneracy of the different model curves largely disappears. In the right panels of figure (6) we show the resolution reached now: we find that the $\chi^2_{TOT}$ rises sharply to a value of $\sim 10^2$ in an interval of $\Delta f_{DM} \sim 0.05$ and of $\Delta \tilde{R}_C < 0.25$. These values set the resolution scale of the fit.\\
\begin{figure}
\begin{center}
\includegraphics[width=10cm,height=20cm]{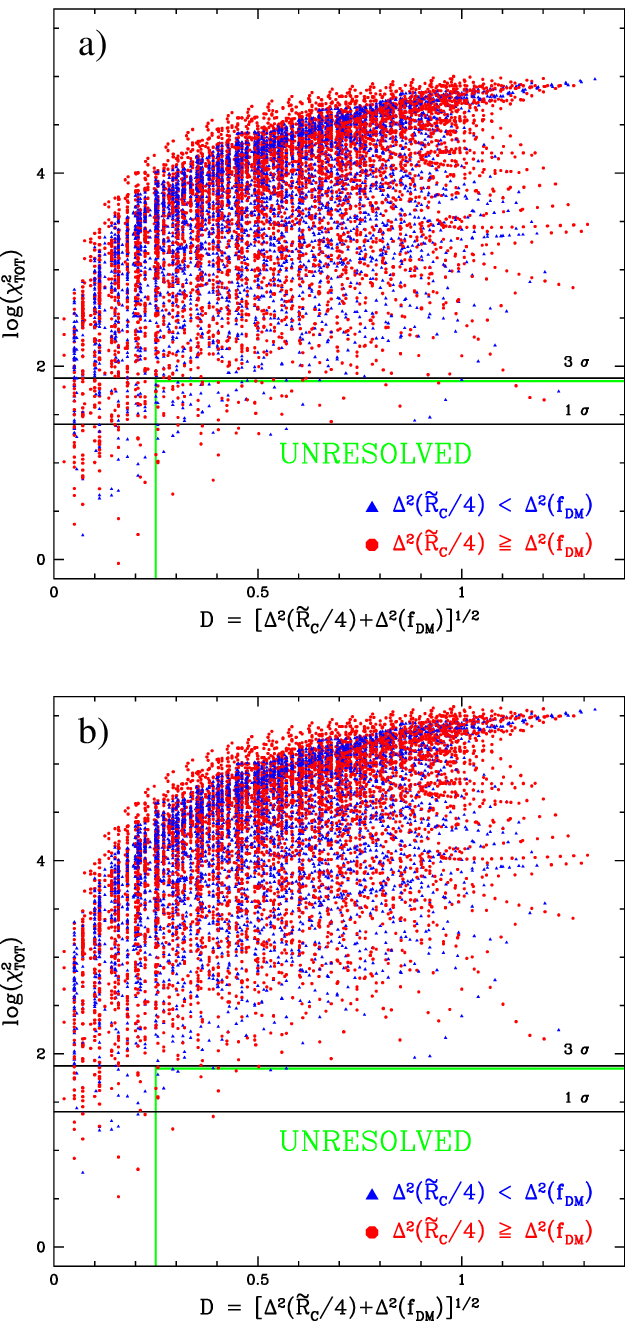}
\caption{$\chi^2_{TOT}$ of the 25 reference ``observed'' curves mapped with 400 model curves, as a function of the distance in the parameter space. Red points (circles) are for distances dominated by the variation of $\tilde{R}_C$, blue points (triangles) are for distances dominated by the variation of $f_{DM}$. The straight lines represent the $1\sigma$ and $3\sigma$ limits. Panel \textit{a)} $\epsilon_V=0.02$ and $\epsilon_D=0.05$. Panel \textit{b)} $\epsilon=0.01$ and $\epsilon_D=0.03$. }
\label{distanze}
\end{center}
\end{figure}
For each one of the 25 ``observed'' reference curves, we perform the fit with the model given by equation (19), considering a very large range for the values of its parameters. We quantify the ``success'' of the fit by the computation of $\chi^2_{TOT}$ for each model curve considered. We now perform a general analysis of the behaviour of $\chi^2_{TOT}$, computed for 10.000 couples \textit{reference ``observed'' RC - model curve}. It is worth to consider $\chi^2_{TOT}$ as a function of the following suitable ``distance'' in the paramater space:
$D \equiv (\Delta^2(\tilde{R}_C/4)+\Delta^2(f_{DM}))^{1/2}$, with $\Delta(x)=x_{mod}-x_{obs}$ (notice that, since the range of variation of $R_C$ is four times that of $f_{DM}$, we normalize the former by a factor of 4 to make the two contributions to $D$ comparable). The result is shown in figure (7a).
The red circles represent cases in which the distance is dominated by the variation of the core radius ( $\Delta^2(\tilde{R}_C/4) > \Delta^2(f_{DM})$), the blue triangles cases in which the distance is mainly due to the variation of the amount of Dark Matter. The straight lines represent the limit value of $\chi^2_{TOT}$ corresponding to $3\sigma$ and $1\sigma$. Most of the points lay over the $3\sigma$ limit, indicating that the two curves of each couple are well resolved and therefore distinguishable.\\
The vertical line defines a distance in the parameter space of $D=0.25$. The points in the range $D\leq0.25$ and laying below the $3\sigma$ limit, although representing cases in which the model fails in fitting uniquely a RC, can be considered irrelevant and will not be counted in the statystics: in fact, the parameters defining the curves are so close to each other, that even if a RC has multiple fits, they differ one from the other by a negligible amount. \\
In the interval $1-3 \sigma$ we find about $0.9\%$ of the total cases, and below the $1\sigma$ limit about $0.23\%$. The cases below the $1\sigma$ limit have to be considered as failures of the model in uniquely distinguishing the Dark Matter distribution, and those below the $3\sigma$ limit as partial failures: therefore, studying our simulated ``observed'' RCs we find that there is a total possibility of only about $1\%$ not to resolve (completely or partially) the mass model.\\
Let's recall that we adopted the observational errors that are typical of about the top $50\%$ of the observed RCs available today, for which, through the application of this procedure, there are very good possibilities to disentangle the mass distribution. However, it is possible to obtain an observed RC with an error even smaller than the one we adopted: for about the top $10\%$ of the RCs available today we can consider observational errors of $\epsilon_V=0.01$ and $\epsilon_D=0.03$, and the number of these curves is going to increase in the future. The same analysis performed with these errors yields a  $\chi^2_{TOT}$ $vs$ $D$ relation as shown in figure (7b). \\
Now we have only about the $0.15\%$ of cases in the interval $1-3\sigma$, and only the $0.02\%$ is below $1\sigma$; the worst case among these sets the lower limit for the resolution of our model: $(\Delta(f_{DM}))_{MAX}=0.25$ and $(\Delta R_C)_{MAX}=1.2R_D$, which is still a reasonable uncertainty.\\
Therefore, considered the very small amount of cases below the $3\sigma$ limit, we can state that the average resolution that we reach corresponds to the step with which we select the model parameters, \textit{i.e.} $(\Delta(f_{DM}))=0.05$ and $(\Delta R_C)=0.2R_D$. \\

\section{Summary and conclusions}

We studied a model galaxy composed of a stellar disk embedded in a dark halo, and we constructed a family of rotation curves by varying the two parameters that affect the RC shape, \textit{i.e.} the fractional contribution of the Dark Matter to the normalized rotation velocity at a reference radius, and the dark halo characteristic scale lenght. We investigated two different profiles for the distribution of the Dark Matter, \textit{i.e.} a pseudo-isothermal and a Navarro, Frenk \& White profile. In both cases we find that no fine-tuning is needed to reproduce the characteristics of the observed galaxy rotation curves, \textit{i.e.} the evidence that they are smooth, with no irregularities like flexes in correspondence of the transition from the disk-dominated into the halo-dominated region. \\
Eventually, irregularities do show up in the PI case, but only for extreme, tuned values of the Dark Matter content and distribution ($f_{DM} \leq 0.1$ and $\tilde{R}_C \geq 6$), and only beyond the disk edge. Since a similar result holds also for the NFW profile, we can conclude that, not only no fine-tuning is needed to reproduce the smoothness of the rotation curves, but the case is the opposite: a fine-tuning is required to obtain a ``marked'' curve. \\
Secondly, with the purpose of investigating how effective is the process of mass modelling the RC, \textit{i.e.} what is the precision with which we can infer a Dark Matter profile from the kinematics, we performed a statistical analysis on a large number of curves, assuming that the Dark Matter distribution is well described by a PI halo. We studied 25 reference rotation curves, representing actual observed RCs, by means of a suitable mass model. We applied a maximum-likelihood method to assign a fitting model curve to each ``observed'' RC, and we found that in the 99\% of cases we resolve the RC and distinguish between different Dark Matter distributions, while in the 1\% of cases we found a degeneracy in the parameters. For this reason we can state that, disposing of observations having a good accuracy and data for a sufficiently high number of radii, the distribution of the Dark Matter inside an observed galaxy can be determined from the RC. \\
Of course, these results refer to the adopted profile used to describe the dark halo, for which the objection of the non-uniqueness of the mass modelling was raised. Also, our results don't minimize at all the importance of complementary methods for determining the mass components of galaxies, like those coming from lensing, or those used to measure the disk masses (see Verheijen 2004).\\
With respect to the more general issue of determining the Dark Matter distribution in galaxies, several different profiles have been proposed to shape the dark halo, and the debate about which one is the favoured is still open. The coming of new-generation telescopes, that will provide high-precision observations of the galactic dynamics, together with an improvement in the resolution of numerical simulations, will surely shed light on this topic. However, a deeper theoretical understanding of the structure of Dark Matter halos may still be required.

\appendix\section{Universal Rotation Curve}
\begin{figure}[h!]
\begin{center}
\includegraphics[width=12cm,height=8cm]{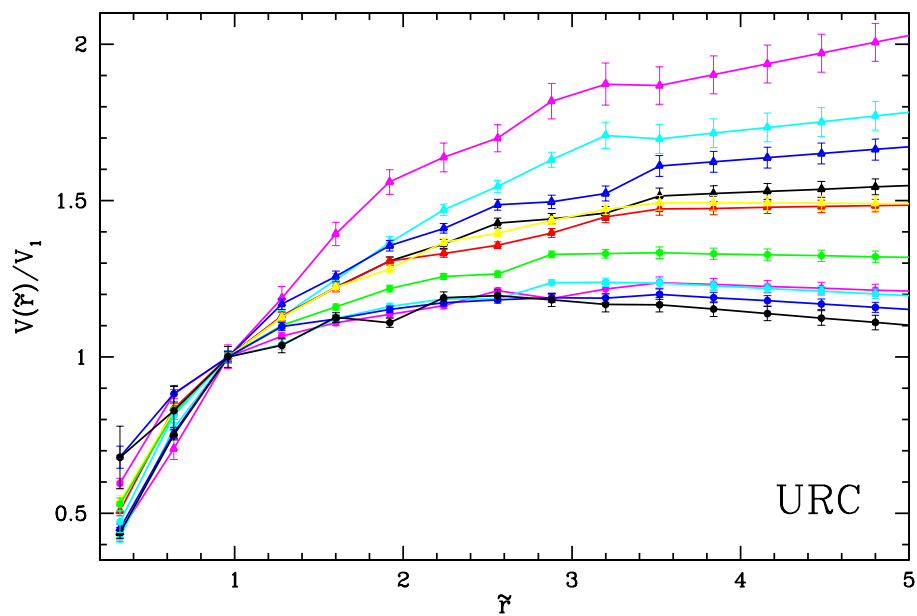}
\caption{The normalized Universal Rotation Curve. See Persic et al.1996 for details.}
\end{center}
\end{figure}

\end{document}